\documentclass{article}
\AtBeginDocument{%
  \providecommand\BibTeX{{%
    \normalfont B\kern-0.5em{\scshape i\kern-0.25em b}\kern-0.8em\TeX}}}

\usepackage[utf8]{inputenc}
\usepackage[english]{babel}

\usepackage{booktabs}
\usepackage{multirow}
\usepackage{comment}
\usepackage{quoting}
\usepackage{subcaption}
\quotingsetup{font={itshape,small},vskip=1pt, leftmargin=15pt}
\usepackage{cleveref}
\usepackage{graphicx}
\usepackage{listings}
\usepackage{tikz}
\usetikzlibrary{arrows, arrows.meta, decorations.pathreplacing,	calligraphy, positioning, fit, calc, backgrounds, decorations.text, external}

\colorlet{black}{black!70!black}
\colorlet{gray}{gray!50}
\definecolor{Peach}{RGB}{255, 224, 178}
\definecolor{Cream}{RGB}{253, 250, 245}
\definecolor{SurgicalGreen}{RGB}{183, 219, 217}
\definecolor{boxGray}{RGB}{127, 127, 127}

\newcommand\myboxRQ[3][]{%
\tikz\node[
  rounded corners=10pt,
  draw=black,
  fill=Peach,
  inner xsep=10pt,
  inner ysep=10pt,
  line width=.4pt,
  align=justify,
  font=\scriptsize,
  text width=\linewidth-20pt-\pgflinewidth,
  #1
]{{\normalsize\textcolor{black}{#2}\smallskip\par}#3\par};}

\newcommand\myboxTaks[3][]{%
\tikz\node[
  rounded corners=10pt,
  draw=black,
  fill=SurgicalGreen,
  inner xsep=10pt,
  inner ysep=10pt,
  line width=.4pt,
  align=justify,
  font=\scriptsize,
  text width=\linewidth-20pt-\pgflinewidth,
  #1
]{{\normalsize\textcolor{black}{#2}\smallskip\par}#3\par};}

\newcommand\myboxHT[3][]{%
\tikz\node[
  rounded corners=10pt,
  draw=black,
  fill=Peach,
  inner xsep=10pt,
  inner ysep=10pt,
  line width=.4pt,
  align=justify,
  font=\scriptsize,
  text width=\linewidth-20pt-\pgflinewidth,
  #1
]{{\normalsize\textcolor{black}{#2}\smallskip\par}#3\par};}

\usepackage{authblk}
\usepackage{cleveref}

\begin{document}

%\title{Designing LLM-Based Voice-Control for Surgical Augmented Reality Navigation System in Pancreatic Surgery}
\title{LLMs Enable Context-Aware Augmented Reality in Surgical Navigation}

\author[1,4]{Hamraz Javaheri}
\author[2]{Omid Ghamarnejad}
\author[1,3]{Paul Lukowicz}
\author[2]{Gregor Alexander Stavrou}
\author[1,3]{Jakob Karolus}

\affil[1]{Corresponding author(s). E-mail(s): Hamraz.Javaheri@dfki.de }

\affil[2]{German Research Center for Artificial Intelligence (DFKI), Germany}
\affil[3]{Klinikum Saarbrucken, Germany}
\affil[4]{RPTU Kaiserslautern Landau, Germany}
\affil[4]{Corresponding author(s). E-mail(s): Hamraz.Javaheri@dfki.de }

%\orcid{0000-0001-5494-0987}

%\orcid{0000-0001-7424-1323}

%\orcid{0000-0003-0320-6656}

%\orcid{http://orcid.org/0000-0002-6890-4271}


\maketitle
\begin{abstract}
Wearable Augmented Reality (AR) technologies are gaining recognition for their potential to transform surgical navigation systems. As these technologies evolve, selecting the right interaction method to control the system becomes crucial. Our work introduces a voice-controlled user interface (VCUI) for surgical AR assistance systems (ARAS), designed for pancreatic surgery, that integrates Large Language Models (LLMs). Employing a mixed-method research approach, we assessed the usability of our LLM-based design in both simulated surgical tasks and during pancreatic surgeries, comparing its performance against conventional VCUI for surgical ARAS using speech commands. Our findings demonstrated the usability of our proposed LLM-based VCUI, yielding a significantly lower task completion time and cognitive workload compared to speech commands. Additionally, qualitative insights from interviews with surgeons aligned with the quantitative data, revealing a strong preference for the LLM-based VCUI. Surgeons emphasized its intuitiveness and highlighted the potential of LLM-based VCUI in expediting decision-making in surgical environments. 
\end{abstract}
\begin{comment}

\begin{CCSXML}
<ccs2012>
   <concept>
       <concept_id>10003120.10003121.10003124.10010392</concept_id>
       <concept_desc>Human-centered computing~Mixed / augmented reality</concept_desc>
       <concept_significance>500</concept_significance>
       </concept>
   <concept>
       <concept_id>10003120.10003121.10011748</concept_id>
       <concept_desc>Human-centered computing~Empirical studies in HCI</concept_desc>
       <concept_significance>500</concept_significance>
       </concept>
 </ccs2012>
\end{CCSXML}

\ccsdesc[500]{Human-centered computing~Mixed / augmented reality}
\ccsdesc[500]{Human-centered computing~Empirical studies in HCI}
\end{comment}
%\keywords{Large Language Models, Voice Control, Surgical Domain, Augmented Reality}



\section{Introduction}
Wearable augmented reality (AR) has become a technology with vast potential for surgical navigation systems, promising enhanced precision and real-time guidance for medical professionals. Despite its promising capabilities, the effective integration of AR in critical domains, such as open surgery, has faced delays and notable challenges compared to other professional fields \cite{edwards2021challenge, fida2018augmented}. One of the challenges associated with this integration delay is the current limitations in usable interaction methods to control the system \cite{saito2020intraoperative, ghazwani2020interaction, coelho2005supporting} that could be easily used and adapted to critical domains \cite{cutolo2019software}. Traditional input mechanisms, such as hand gestures or using combinations of eye gaze and virtual menus, prove impractical in areas where manual control is restricted and the AR view should not be occluded with too many virtual objects, hindering the full adaptation and utilization of wearable AR technology in these crucial fields \cite{mentis2015voice, wachs2011vision}.

Among alternative interaction methods, voice-controlled assistants using speech commands stand out as a viable option, offering a hands-free and potentially intuitive means of interacting with AR systems during surgical procedures \cite{mentis2015voice, blokvsa2017design, hertel2021taxonomy, hatscher2018hand}. While impressive in their ability to respond to voice commands, often lack contextual understanding and adaptability \cite{gowthamy2023enhanced}. Furthermore, Relying solely on speech commands presents its own set of challenges, including the need to implement distinct keywords for each custom functionality, consequently increasing the complexity of the application for users \cite{mentis2015voice}. In domains where simplicity, efficiency, and ease of use are mandatory, such as in critical surgical settings, introducing additional workloads or time-consuming processes may compromise the technology's adoption and effectiveness.

Despite recent improvements in interaction methods for AR applications \cite{nizam2018review} and in voice-controlled assistant systems\cite{dutsinma2022systematic}, a gap still exists in the feasibility of using these techniques in critical domains, such as surgical navigation systems \cite{cutolo2019software, saito2020intraoperative}. The unique challenges presented in such vital domains have not yet received sufficient attention. 

In this paper we presented a voice-controlled user interface (VCUI) for a surgical AR-based assistance systems (ARAS), utilizing large language models (LLM), aiming to address the aforementioned limitations and enhance user experience in these critical settings. By harnessing the capabilities of speech recognition and LLMs, we introduce an intuitive method to control the system during surgical procedures helping to reduce the cogntitive load caused by interaction with the system. 

We use LLM to perform function calls and control the system, enabling users to execute parallel functionalities based on the context of their request and the context presented in the application. We integrated our LLM-based VCUI into the previously developed and clinically evaluated system, ARAS, designed to navigate pancreatic surgeries by enabling in-situ visualization of patient-specific 3D models of vessels and tumors, as well as access to supportive data.

We evaluated the feasibility of our LLM-based VCUI for intraoperative interaction with ARAS throughout two studies, comparing it against previously tested VCUIs for surgical AR applications using speech commands  \cite{scherl2023augmented, dias2021augmented}. 
Firstly, we conducted a user study with expert surgeons (N=9) involving surgical tasks to simulate the environment where ARAS would be normally used. We measured system usability, task execution time (TCT), cognitive load (NASA-RTLX), and conducted interviews to evaluate our method through quantitative and qualitative measures. 
Secondly, we employed both VCUIs independently in two pancreatic surgeries to further evaluate these approaches in the users' end setup.
Our findings highlight that the integration of LLM-based VCUI led to faster task executions and reduced cognitive workload by generating more context-aware system outputs compared to VCUI relied on speech commands.

%Our work contributes to the design and development of voice-based interaction systems specifically for time-critical and demanding domains, such as surgery. In this study, we focused on evaluating our proposed LLM-based voice interaction method and comparing it with state-of-the-art voice interaction methods, specifically speech commands for interacting with an AR-based surgical assistance system.
%The evaluation of the AR-based surgical assistance system falls outside the scope of this paper. 

Our work contributes to the design and development of VCUIs specifically for time-critical, highly stressful, and demanding domains, such as surgery. These domains often do not fit general solutions but require systems that are carefully designed, implemented, and include user involvement throughout the process, as any unpredicted errors during system implementation could have significant consequences. To the best of our knowledge, our work is among the first studies to explore the potential of controlled usage of LLMs targeted and tested in a surgical environment beyond their commonly explored role as training and teaching tools \cite{mohapatra2023leveraging, varas2023innovations}. Our study showcases the feasibility of this approach in such a critical field by demonstrating not only the advantages of LLMs for users in terms of time and cognitive load reduction but also how LLMs can simplify the development process by combining and calling simple system functionalities to achieve more complex and context-aware system behavior.

\section{Related Work}
This related work section will first provide a comprehensive overview of AR applications in the surgical domain, followed by an exploration of interaction techniques within the medical field, and conclude with a dedicated chapter on natural communication and speech-Based interaction.

\subsection{Augmented Reality in Surgery}
Surgical assistance and simulation systems have undergone significant advancements with the integration of AR technologies \cite{vavra2017recent,zhang2019preliminary, lungu2021review}. While many approaches were implemented for display-based or handheld AR navigation systems for surgical procedures for example in laparoscopic surgery \cite{qian2019review}, the advancement of having wearable AR devices opens the door for further integration of technology in more challenging fields such as open surgery \cite{fida2018augmented,pratt2018through, scherl2021augmented}. While the interaction principles to execute certain tasks with displays and touch-sensitive surfaces such as tablets have become traditional and well-accepted over the years, the control techniques for wearable AR devices have yet not been perfected. The restrictions available in critical fields such as open surgery in terms of available interaction and control methods, make the integration of this technology in a routine practice more difficult as the risks are very high and any distractions or errors caused by technological devices are not welcomed. In a study done by Saito et al. \cite{saito2020intraoperative} it was demonstrated that using 3D holograms and hand manipulation caused higher cognitive workload compared to usual 2D supports as it required higher physical demand and effort. 

\subsection{Interaction Techniques in Medical Domain}
The interaction techniques used in wearable AR technologies might vary depending on the task and used modalities \cite{hertel2021taxonomy}. While many interaction modalities could be used for a range of tasks in non-critical medical domains, such as training and simulations\cite{allgaier2022comparison,khundam2021comparative}; the options for hygienic and sterilized surgical environments remain very restricted. Among different interaction and control modalities hand, voice, and foot input became possible options for interactions in the surgical environment as they do not require any direct contact with foreign objects \cite{hatscher2018hand}. Despite, the usability of these interaction methods, the environmental limitations during the surgical theater might not always allow practical usage of all these input modalities \cite{mentis2015voice, wachs2011vision}. Moreover, the limited number of different gestures that could be performed using only hand or foot inputs might restrict the system features \cite{hertel2021taxonomy} when compared to voice. 

\subsection{Natural Communication and Speech Based Interaction}
The employment of voice input as a natural interaction, control, and communication method gathered extensive attention in the research. While the majority of applications focused on using voice in combination with other input modalities \cite{ismail2015multimodal,nizam2018review,lee2013usability}, the recent speech recognition algorithms and natural language processing provided mediums to use voice-based interaction as the sole input modality \cite{terzopoulos2020voice}. With the latest development of smart assistant systems and natural communication schemes through speech, voice input became popular especially where other input modalities could not be used \cite{hoy2018alexa}. With the outbreak of LLM, new possibilities have emerged to use natural communication schemes for interaction with assistant systems. Mahmood et al. \cite{mahmood2023llm} presented a LLM-powered conversational voice assistant that could be used in different areas. The combination of speech input and LLMs could be also used to achieve smarter assistant systems to control the system and perform certain system functionalities, as it was demonstrated by Dong et al.\cite{dong2023towards}. 

Despite the extensive research in the mentioned research fields, there has been a notable gap in exploring the various VC methods within AR environments. This gap in the literature is particularly noteworthy in critical domains such as open surgery, where practicality and safety become essential, and alternative input modalities may not be feasible.

\section{Methodology}

\begin{figure}[htbp]
    \centering
    \includegraphics[width=\columnwidth]{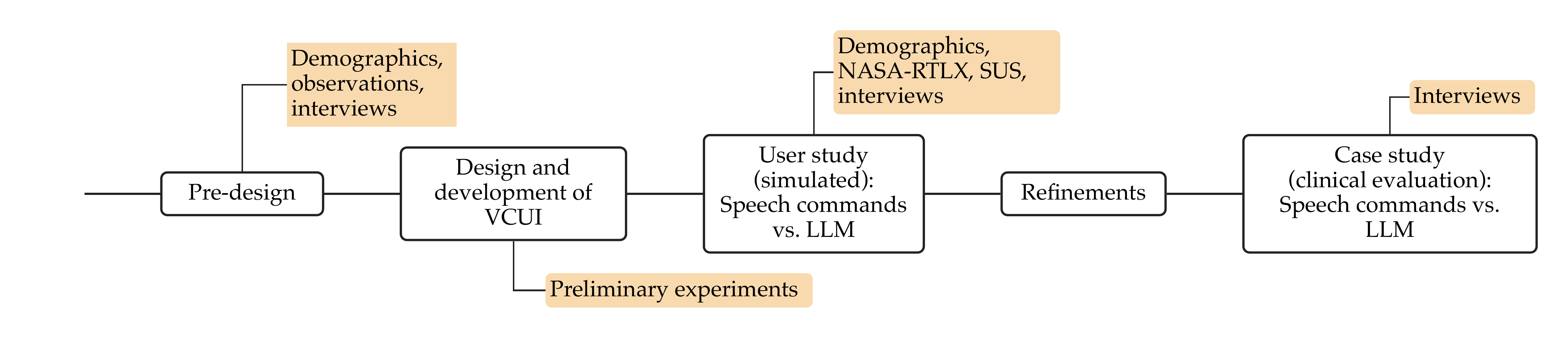}
    \caption{Chart showing different phases of our work from pre-design till the case study involving a clinical trial and associated data collection for each phase.}
    \label{fig:procedure}
\end{figure}

In this work we focused on finding an optimal interaction method for ARAS specifically designed for open pancreatic surgery, addressing the challenges associated with the impracticality of common input modalities in confined surgical spaces. This led us to explore and evaluate voice control (VC) methods for a more practical and efficient user experience in this surgical context.
To guide our research, we addressed the following research questions over five consecutive phases as depicted in \Cref{fig:procedure}. 

\vspace{10pt}
\begin{frame}{}
\begin{minipage}{\linewidth}
  \myboxRQ{\textbf{Research Questions (RQs):}}{
  \normalsize
  \begin{enumerate}
      \item What are the user and field-specific requirements in terms of interaction methods for an AR-based surgical navigation system?   
    \vspace{5pt}
    
     \textit{\textbf{Objective:}} To gather insights from surgeons to inform the design of the user interface and interaction with the AR system, ensuring it meets their practical requirements and enhances their ability to perform the surgery.
        
    \textit{\textbf{Method:}} Observations, interviews with surgeons, along with demographics of participating surgeons.

    \vspace{5pt}
     \item How feasible is an LLM-based VCUI for a surgical AR system, and how does it impact the user’s cognitive load? How does this approach compare to previously tested VC methods, such as speech commands?
    \vspace{5pt}
 
      \textit{\textbf{Objective:}} To gather insights from surgeons about the usability, cognitive workload, and their assessment of the LLM-based VCUI method and compare it to the previously employed approaches.
        
    \textit{\textbf{Method:}} User study with surgeons in a simulated scenario involving surgically relevant system interaction tasks. Data collection using NASA\_RTLX~\cite{hart2006nasa}, and system usability scale (SUS)~\cite{brooke1996sus}, and post-study interviews with surgeons, along with demographics of participating surgeons. 

     \vspace{5pt}
    \item How feasible is the LLM-based VCUI in the users' end setup during surgery compared to speech commands? What are the users' (surgeons') reflections?
     \vspace{5pt}
     
     \textit{\textbf{Objective:}} To gather insights from field surgeons about the usability of each VCUI in an ecologically valid setup that involves highly stressful situations.
        
    \textit{\textbf{Method:}} Case study involving the employment of both VCUIs, each in a pancreatic surgery session, and conducting postoperative interviews with surgeons about the interaction method used to control the surgical AR system.
    \vspace{2pt}
    \end{enumerate}

}
\end{minipage}
    
\end{frame}  
\vspace{10pt}
We began our investigation by exploring the user and domain-specific requirements for a VCUI for ARAS by interviewing experts from the field. Consequently, we developed two non-conversational VCUI.
Following the previous works on clinically tested voice interaction methods, our first VCUI utilizes speech recognition and speech commands \cite{scherl2023augmented, dias2021augmented} and serves as our baseline. The second VCUI incorporates speech recognition, an LLM, and natural communication schemes to control the system.  
Internally, both VCUIs have access to the same set of system functions of ARAS. 
In our first user study, we explored the usability of each VCUI in a simulated environment and compared their performance in terms of added workload on surgeons while performing surgically relevant tasks, their usability, and conducted semi-structured interviews with participating surgeons.
Finally, after proving the usability of both methods, we tested both VCUIs in an ecologically valid environment during a clinical trial and conducted interviews with surgeons.

In this work, we solely focused on evaluating our proposed LLM-based VCUI for ARAS in a time-critical domain, specifically in pancreatic surgery, and compared our approach to a conventional VCUI using speech commands. We note that the evaluation of ARAS itself falls outside the scope of this paper.

%To do so we started the process with interviewing four surgeons from Hepatopancreatobiliary and visceral field (\Cref{DesignInterviewParticipants}) to understand the user requirements and environmental challenges associated with interaction with AR navigational system in surgical domain. 

\section{System Design and Development}

We used the Unity 3D game engine \cite{unity3d} for software design and development and used Microsoft HoloLens 2 device \cite{hololens2} as a wearable AR device.
The selection of the HoloLens 2 for this application was mainly motivated by ethical, and safety considerations as it is CE-certified and was shown to have been successfully used in the medical domain before~\cite{ivanov2022practical, liu2021augmented, saito2020intraoperative}.

\subsection{AR Assistance System}
\begin{figure}[htbp]
    \centering
    \includegraphics[width=\columnwidth]{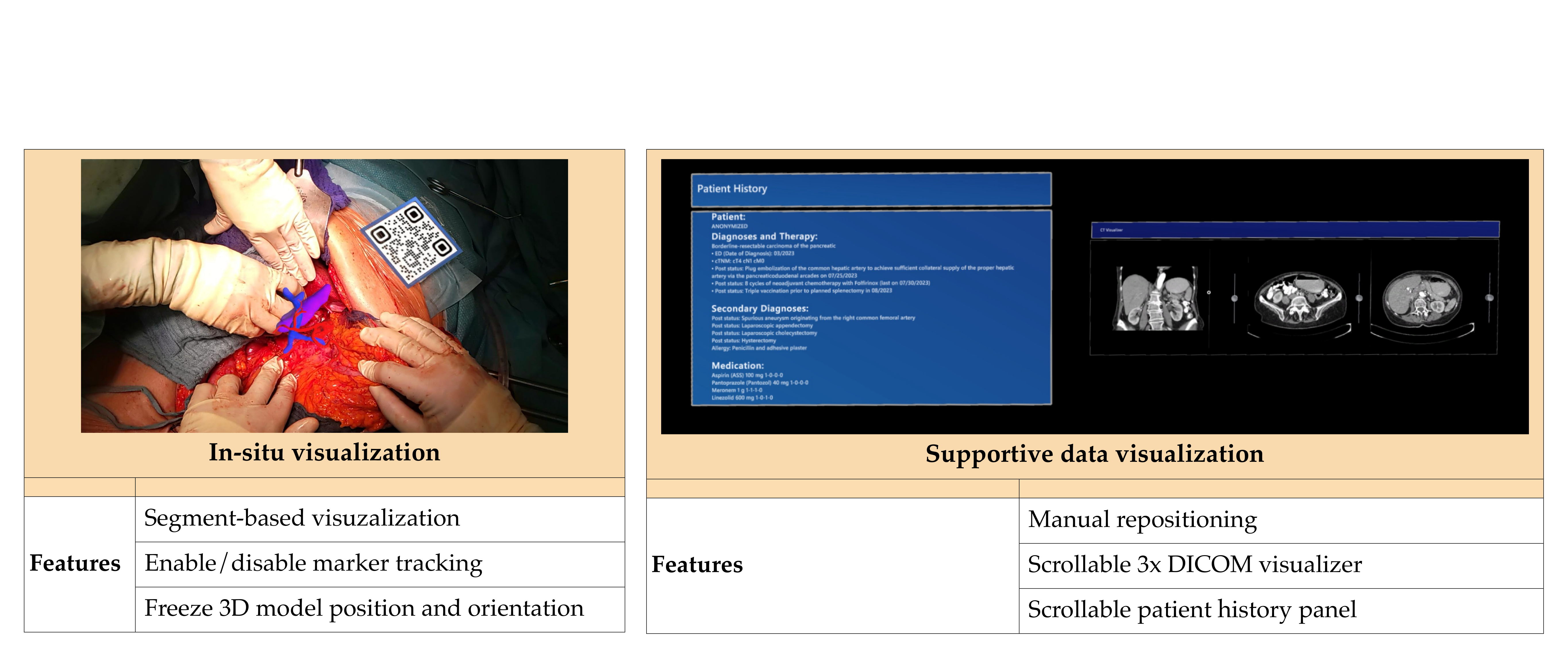}
    \caption{The characteristics of in-situ visualization and supportive data visualization feature set provided by the ARAS software.}
    \label{fig:ARAS}
\end{figure}

The ARAS software was designed and clinically evaluated as a supportive and navigation tool for open pancreatic surgery. It was designed to contain two distinct feature sets to support surgeons throughout the surgery session: in-situ visualization and supportive data visualization (\Cref{fig:ARAS}). The system features segment-based visualization of a patient-specific 3D model, the ability to enable or disable modes between marker-based tracking for in-situ visualization, and maintaining the model's position and orientation. It also allowed for the visualization of supportive data, such as CT images and patient diagnoses. The designed user interface was aimed to provide means for surgeons to precisely execute features without interrupting the surgical flow.

\subsection{Voice Control User Interfaces}
\begin{figure}[htbp]
    \centering
    \includegraphics[width=0.6\columnwidth]{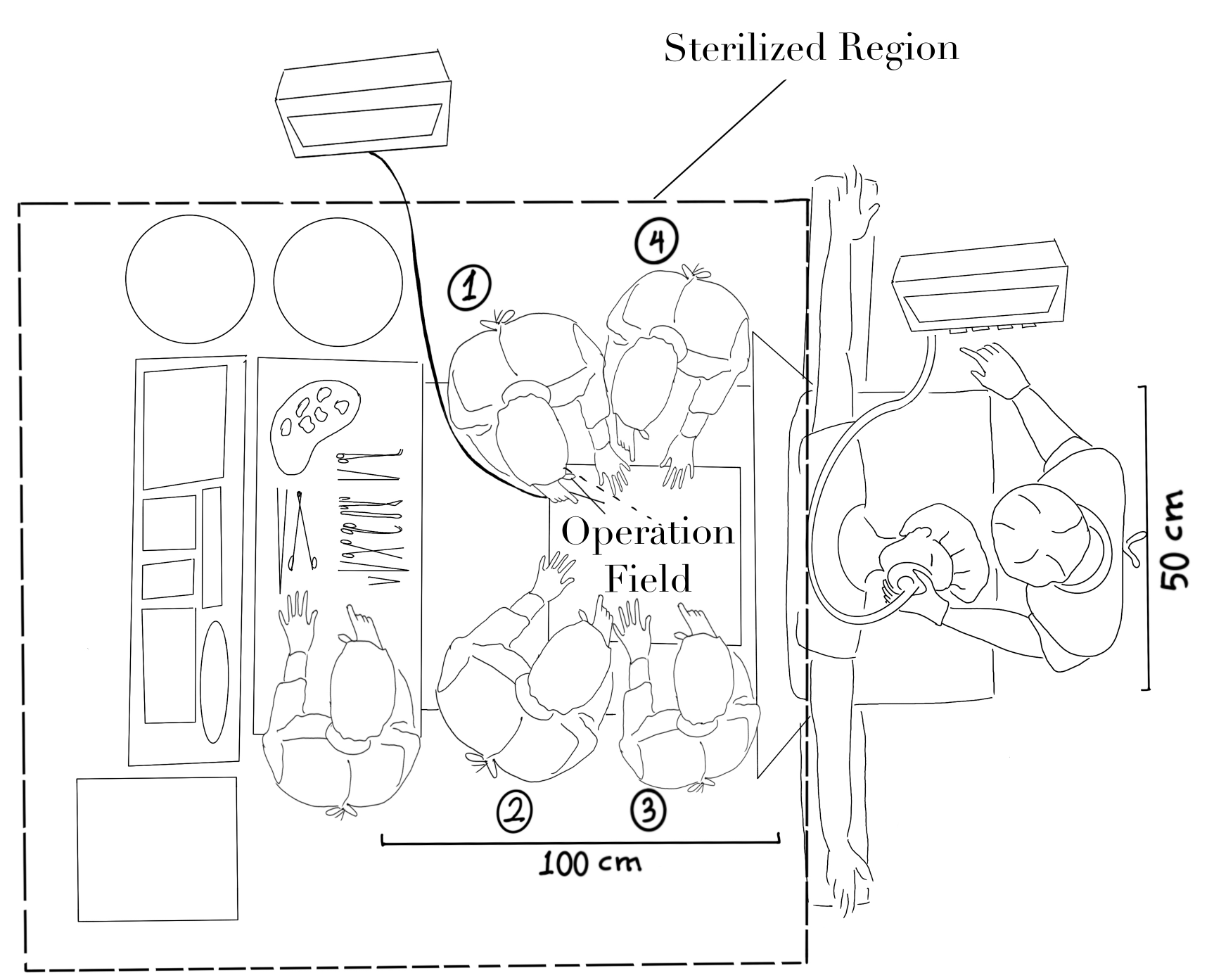}
    \caption{A top view of the surgical room setup. The placement of the medical staff around the table. The limited area around the operation field, and sterilization rules restrict the interaction with AR surgical assistance system. The surgeons are numbered based on their role in the surgery with Number 1 being the lead surgeon.}
    \label{fig:surgeryTable}
\end{figure}

During the pre-design stage, we interviewed four surgeons from the Hepatopancreatobiliary \footnote{ Hepatopancreatobiliary surgery consists of the general surgical treatment for benign and malignant diseases of the liver, pancreas, gallbladder, and bile ducts.} and Visceral\footnote{Visceral surgery, also known as abdominal surgery, refers to surgery of the abdominal cavity and abdominal wall, endocrine glands, and soft tissue, including transplantation.} fields (\Cref{tab:DesignInterviewParticipants}) to understand the user and domain-specific requirements in terms of interaction with ARAS. Furthermore, to enhance our understanding of the domain-specific constraints and environmental challenges we participated in 2 pancreatic tumor resection procedures as observers. The insights from interviews and observations made during the operation highlighted certain limitations regarding the feasible interaction method to control ARAS. With surgeons requiring both hands to perform intricate operations, incorporating hand gestures is severely constrained. Furthermore, the physical setup around the operation field of the surgery table, where four surgeons and a nurse typically stand (as illustrated in \Cref{fig:surgeryTable}), results in a limited field of view for the AR application. Consequently, placing any virtual objects between surgeons and the operation area can obstruct the surgical view, introducing undesirable visual occlusion. Objects positioned behind other medical professionals are not only obscured but also challenging to interact with, given the physical constraints of extending hands and arms in a confined space. These insights underscore the necessity of an alternative user interface, leading us to explore and evaluate fully VCUI for a more practical and efficient user experience in the context of surgery. Consequently, we developed two VCUIs to interact with ARAS. While the first VCUI follows the previously tested VC methods for the surgical domain using speech commands, our second VCUI uses LLM and a natural communication scheme.

\begin{table}[htbp]
\centering
\caption{Profiles of interviewed surgeons before design and development process. }
\begin{tabular}{llll}
\toprule
\multirow{2}{*}{ \textbf{Age}}  & \multirow{2}{*}{ \textbf{Gender}} & \multirow{2}{*}{ \textbf{Field}} & \textbf{Surgical}\\
& & & \textbf{Experience (year)}  \\
\midrule
51 & Male & Hepatopancreatobiliary & 23 \\
53 & Male & Hepatopancreatobiliary  & 25    \\
40 & Female &  Hepatopancreatobiliary  & 13   \\
33 & Male & Visceral  & 4  \\

\bottomrule
\end{tabular}
\label{tab:DesignInterviewParticipants}
\end{table} 

\subsubsection{VCUI with Speech Commands}

\begin{figure}[htbp]
    \centering
    \includegraphics[width=0.6\columnwidth]{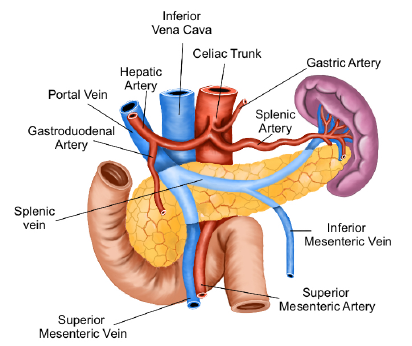}
    \caption{Anatomical drawing of 3D reconstructed segments in ARAS and their positions around the pancreas. The names of these segments were used as keywords for VCUI using speech commands. }
    \label{fig:segment}
\end{figure}

In implementing the speech commands, we utilized the Windows speech recognition service (using IMixedRealityDictationSystem service) \cite{Microsoft2020WindowsSpeechInputProvider} and Microsoft Mixed Reality Toolkit, MRTK2 \cite{Microsoft2022MRTKUnity}, to define specific keywords for triggering desired functionalities. These keywords were carefully selected in collaboration with surgeons from the field, ensuring relevance and ease of recall. A total of 34 unique keywords were assigned to trigger the system functionalities (\Cref{tab:VCUI}). Designed as medical terms commonly used in daily practice or intuitive words, each keyword corresponded either to the names of anatomical structures available in the 3D model (\Cref{fig:segment}) or as intuitive words routinely used in a clinical environment or daily life. Recognizing the challenges of keyword recognition and aiming to enhance flexibility, some functions could be executed using multiple keywords or synonyms. For instance, the keyword "cancer" could be used interchangeably with the keyword "tumor" to enable or disable tumor visualization (\Cref{tab:VCUI}). Additionally, all the keywords used for 3D visualization could be used in a combination of ON/OFF to either enable or disable some segments visualization or without these extensions as toggle behavior between on and off mode. 
To provide visual feedback to the user confirming the successful recognition of the spoken keyword, we used the MRTK tooltip component to be shown upon the detection of a speech command.

\begin{table}[htbp]
\centering
\caption{Implemented voice keywords and their functionalities}
\resizebox{\textwidth}{!}{%
\begin{tabular}{lll}
\toprule
\textbf{Type} & \textbf{Voice Keywords} & \textbf{Functionality}\\
\midrule

\multirow{2}{*}{Patient information} & Patient history / diagnosis / medication & Activates/Deactivates patient history panel \\
& CT / CT Image / Tomography /& \multirow{2}*{}{Activates/Deactivates CT Visualizer }\\
&  Computed Tomography / CT scans \\

\midrule 
\multirow{8}{*}{3D visualization} & arteries/veins & Activates/Deactivates rendering of all arteries/veins \\

& Sternum & \multirow{8}{*}{Activates/Deactivates rendering of the associated structure}\\
& Celiac trunk &    \\
& GDA  &  \\
& Mesenteric artery &  \\
& Splenic artery &  \\
& Gastric artery &  \\
& Hepatic artery / liver artery &  \\
& Portal vein &  \\
& Vena cava &  \\
& Splenic vein &  \\
& Mesenteric vein &  \\
& Tumor / lesion / cancer &  \\

& Variation & Activates/Deactivates rendering of unusual anatomical structures \\
\midrule
\multirow{3}{*}{Control commands } & Go up/down & Activates the automatic scrolling up/down of the CT slices \\
 & Stop  & Stops the automatic scrolling of CT slices \\
 & Capture photo / hologram & captures a photo using HoloLens mounted camera with / without holograms\\

& Freeze & Freezes the 3D model position and orientation and disabling the marker tracking \\
& Marker tracking & Enables marker tracking \\
& Reset & Resets the app including the 3D model position and orientation relative to sternum marker\\

\bottomrule
\end{tabular}
}
\label{tab:VCUI}
\end{table}

\subsubsection{LLM Based VCUI}
\begin{figure}[htbp]
    \centering
    \includegraphics[width=\textwidth]{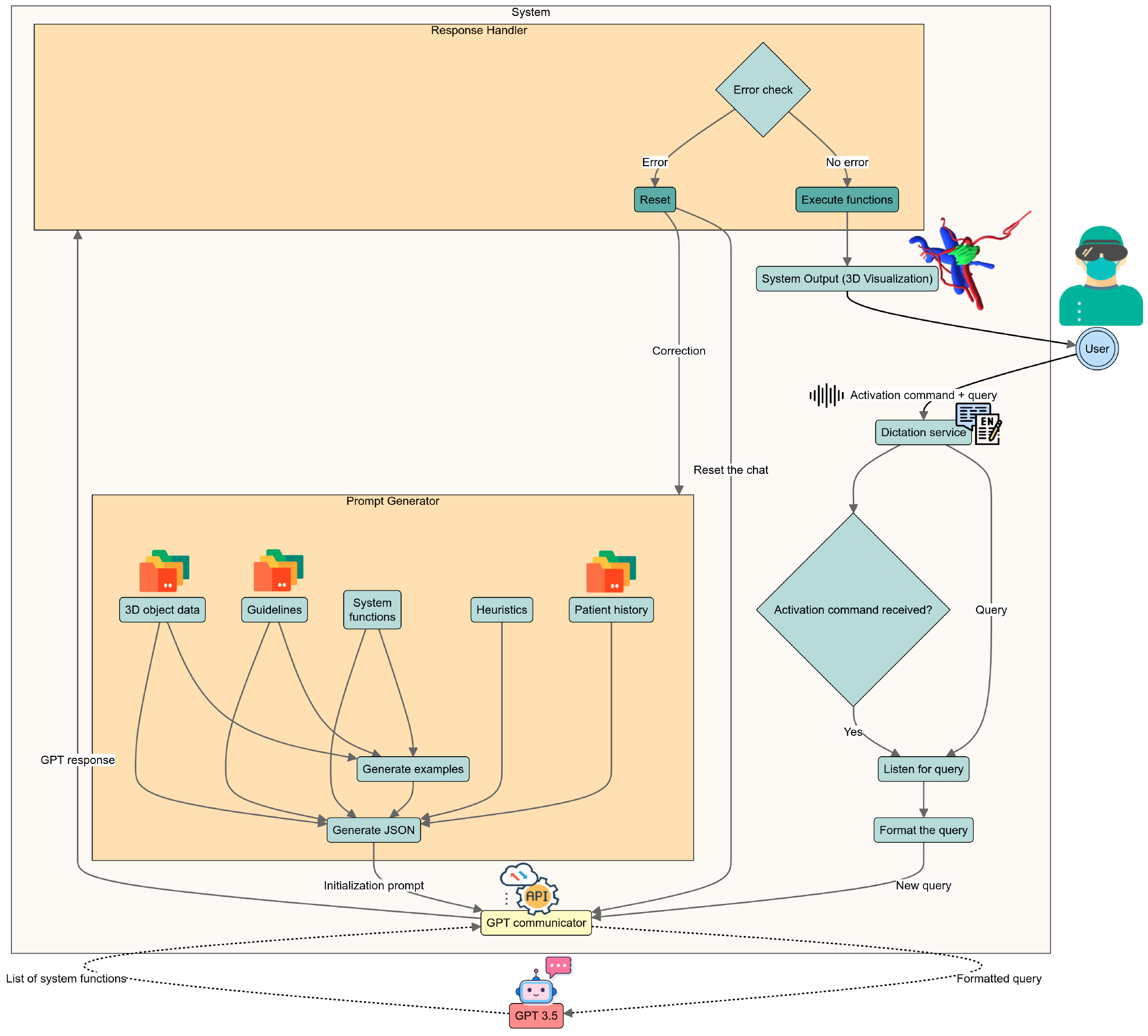}
    \caption{Overview of LLM-based VC framework. The system begins with loading patient files and function descriptions to generate an initial prompt. The system functions then is called upon the user's query via speech. }
    \label{fig:systemArchitecture}
\end{figure}

We designed and implemented a system framework to automatically execute system functions upon user query stated via natural speech and speech recognition using the same Windows speech recognition service as the speech commands method \cite{Microsoft2022MRTKUnity}. We implemented a GPT communicator layer for Unity in C\# to facilitate external API calls to chat-GPT from Unity and used GPT-3.5-turbo model to process the user request and return the system functions.
Our LLM-based VCUI framework consisted of four main components, a dynamic initial prompt generator, a dictation service, a response handler, and a GPT 3.5 model (\Cref{fig:systemArchitecture}). 

%On initialization, the system generates an initial prompt and sends it to the GPT model. 

We developed an adaptive prompt generator to dynamically create initial prompts specific to each patient, aiming to provide patient-specific contextual information while defining the task for the LLM.

We used the patient's specific 3D model meshes to calculate the proximity and relational distance of each 3D object in the model, such as vessels and tumors, to other structures. We further provided the system with the patient diagnosis, and surgical resection guidelines, along with a list of system functions, and heuristic examples (\Cref{fig:systemArchitecture}). %The reason for providing all this information was to give the LLM more context about the patient's case, specifically regarding the observed 3D model, enabling the LLM to provide more case-specific outputs based on the patient.

To mitigate hallucinations by LLMs, as demonstrated in other related studies \cite{yao2023llm, leiser2024hill, galitsky2023truth}, we implemented an auto-repeat function that resends the initial prompt to the chat. Our preliminary experiments revealed that the accuracy of LLM outputs decreases as more user inputs are added, causing the model to lose the context of the initial prompt—which contains critical patient-specific information and the main task. This leads to increased hallucination in the generated responses.

To counter this effect, we developed a mechanism that periodically reminds the LLM of the task and relevant information by resending the initial prompt after each user input. Once the user's request is processed, this reminder prompt is automatically sent to the chat, ensuring that the model retains the original context without the user noticing.

As a safeguard for occurred hallucinations which resulted in non-accurate response from LLM, we have implemented a reset function that would be used to correct the initial prompt and reset the chat. Like other system functionalities, the LLM could call upon this function based on the context of the user query. Users were informed about this functionality and instructed that if the system executed an incorrect or unexpected action, they could notify the LLM and specify the correct response for that situation.
 Apart from this reset function, no extra function or direct annotation to the study tasks (C\ref{study_design}) was included in the initial prompt to avoid potential performance bias for the sake of the study. 

As the GPT model only receives data in text format, we generated a JSON file containing all this information along with the requested task to return the appropriate system functions and variables based on the given sentence. The JSON format used for the initial prompt is given in \Cref{appendix_json}.

\begin{comment}

\begin{figure*}[t]
    \centering
    \includegraphics[width=\textwidth]{Figures/Initialization Prompt.png}
    \caption{The flowchart of the dynamic initial prompt generator system that adapts the initial prompt for each patient case using patient files, guidelines, and heuristic examples.}
    \label{fig:initializationPrompt}
\end{figure*}

\end{comment}
After successful initialization, the user could send a request to the application through voice query using a natural communication scheme. The dictation is activated using a single speech command called "Assistant" to avoid false queries. After activation of the dictation system, the system starts listening to the user query. During preliminary testing, we observed that most of the requests from users to the LLM last around ten seconds. Therefore, we set the default listening time to ten seconds. However, if the ten seconds is exceeded and the user is still speaking the system would wait for two seconds of silence before sending the query to the LLM. This way we made sure that the system would at least listen to the user query for ten seconds while also allowing for longer queries.
 The transcribed user query would then be formatted to JSON and sent to the GPT model via the GPT communicator. 
 Upon receiving the response from GPT, the response is first validated, and if there is no error in the received format or request to reset the chat by LLM due to the user correction, then associated system functions are executed and presented to the user. 
 
If there is a call to reset among the received functions from LLM, then the system stores the user interaction example along with the correction that the user provided to LLM. The active chat would be terminated another initial prompt would be generated using the recently added user example and a new chat would be initiated.

In addition, similar to tooltips used in the speech commands approach, we designed a virtual panel (\Cref{fig:manikin}) to provide a real-time transcription of the user's voice input, to provide visual feedback about the recognition of the user's query. This decision was made to show the user about the recognized request and give the user a chance to correct it if it was detected wrong.

 \section{User Study 1: Evaluation During Simulated Surgical Scenarios} \label{sec: study1}

We employed a mixed-method evaluation approach, integrating both qualitative and quantitative data analyses, to comprehensively assess the feasibility of our proposed LLM-based VCUI and draw comparisons with the VCUI method utilizing speech commands. In this study, we compared and evaluated both methods in a simulated lab environment focusing on our two research questions (RQ2). We conducted a within-subject study, where all participants experienced both VCUI (LLM and speech commands) for two different patient cases. To avoid potential bias the order of the VCUI and the patient case were counterbalanced. 

\subsection{Study Design}
\label{study_design}
Aligned with ARAS’s primary function, the study tasks focused on adjusting the visualization of 3D model segmentations to guide various phases of surgery. This approach aimed to test the VCUIs within their relevant context in a controlled, simulated environment.

%, as ARAS was designed to provide intraoperative visualization of patient-specific 3D models which includes all non-visible structures hidden under tissue and organs to guide surgeons during pancreatic resections, a complex procedure involving intricate anatomical considerations. 

Even though the tested system's functionality was limited, this task design aimed to focus more on interaction with the system as a result of cognitively challenging tasks which would be the normal case during the pancreatic surgery. The decision to trigger different system functions to visualize various combinations of structures depended on several factors, such as the relationship between structures and the tumor, patient history, surgical guidelines, and the current stage of the surgery. Any unnecessary virtual objects or structures visible in the AR view could confuse the surgeon or unnecessarily occlude the view.
It is worth noting that while it is possible to implement separate functions for certain predefined guidelines, often the decision of which structure combination to visualize is not definite, hard to implement, requires processing power on the operating device, and highly varies depending on each patient case and the stage of the operation. 

Given the multifaceted nature of pancreatic surgery, which is divided into consecutive sessions for the preparation and resection of the vascular system and pancreas organ infiltrated with the tumor, we tailored the tasks to align with the interaction with the system during these distinct stages of each intraoperative session.

The task designs and their execution order were advised by experienced surgeons to simulate the progression of pancreatic surgery and the associated workload on surgeons. This approach aimed to replicate the decision-making process, considering the varying cognitive difficulty levels involved in identifying vital structures at different stages of the surgery.

The first two tasks targeted the commonly used first surgical approaches therefore the names of the structures to be visualized were given in the task description. Tasks 3 and 4 are performed in occasional situations based on the progress of the operation, however, due to the fix procedure performed in these tasks the names of the essential structures to be visualized were also given in the task description. Tasks 5 and 6 were designed to address later stages of the operation where the surgeon is required to make complex decisions on which structures are needed to be observed to guide a critical phase of the operation such as the tumor resection phase. Therefore, in these last two tasks, the names of the structures to be visualized were neither given in the task description nor were annotated in the system as the decision to which structure to be visualized is subjective to the surgeon and might vary. While task 5 focuses on visualization of the structures that are affected by the tumor, task 6 focuses on visualization of the structures that need to be removed along with the tumor which is not always limited to those structures affected by the tumor. The task descriptions were as follows:

\vspace{10pt}
\begin{frame}{}
\begin{minipage}{\linewidth}
  \myboxTaks{\textbf{Tasks:}}{
  \normalsize
  \begin{enumerate}
     \item \textbf{Kocher maneuver:} During this task participants were asked to only enable the visualization for the following structures: the tumor, inferior vena cava, and portal vein \cite{venes2017taber}.

    \item \textbf{Preparation of the hepatoduodenal ligament: } During this task participants were asked to enable visualization for the following structures: portal vein, hepatic artery, and gastroduodenal artery \cite{venes2017taber}.

    \item \textbf{Uncinate-first approach:} During this task participants were asked to enable visualization for the following structures: tumor, portal vein, and superior mesenteric artery \cite{venes2017taber}.

    \item \textbf{Artery-first approach:} During this task participants were asked to enable visualization for the following structures: hepatic artery, gastroduodenal artery, celiac trunk, and superior mesenteric artery \cite{venes2017taber}.

    \item \textbf{Tumor infiltration: } 
    Participants were asked to only enable those structures that are infiltrated by the tumor. They were free to look at the CT images to decide or observe the 3D model as it would be the case for the real surgery.

    \item \textbf{Complex surgical decision: } During this task the participants were asked to evaluate and make decisions about which structures should be resected (surgically cut or removed along with the tumor to secure patient safety) during the pancreatic resection and only enable the visualization of those structures.
    
    \end{enumerate}

}
\end{minipage}
    
\end{frame}  
\vspace{10pt}

\begin{figure}[htbp]
    \centering
    \includegraphics[width=0.7\columnwidth]{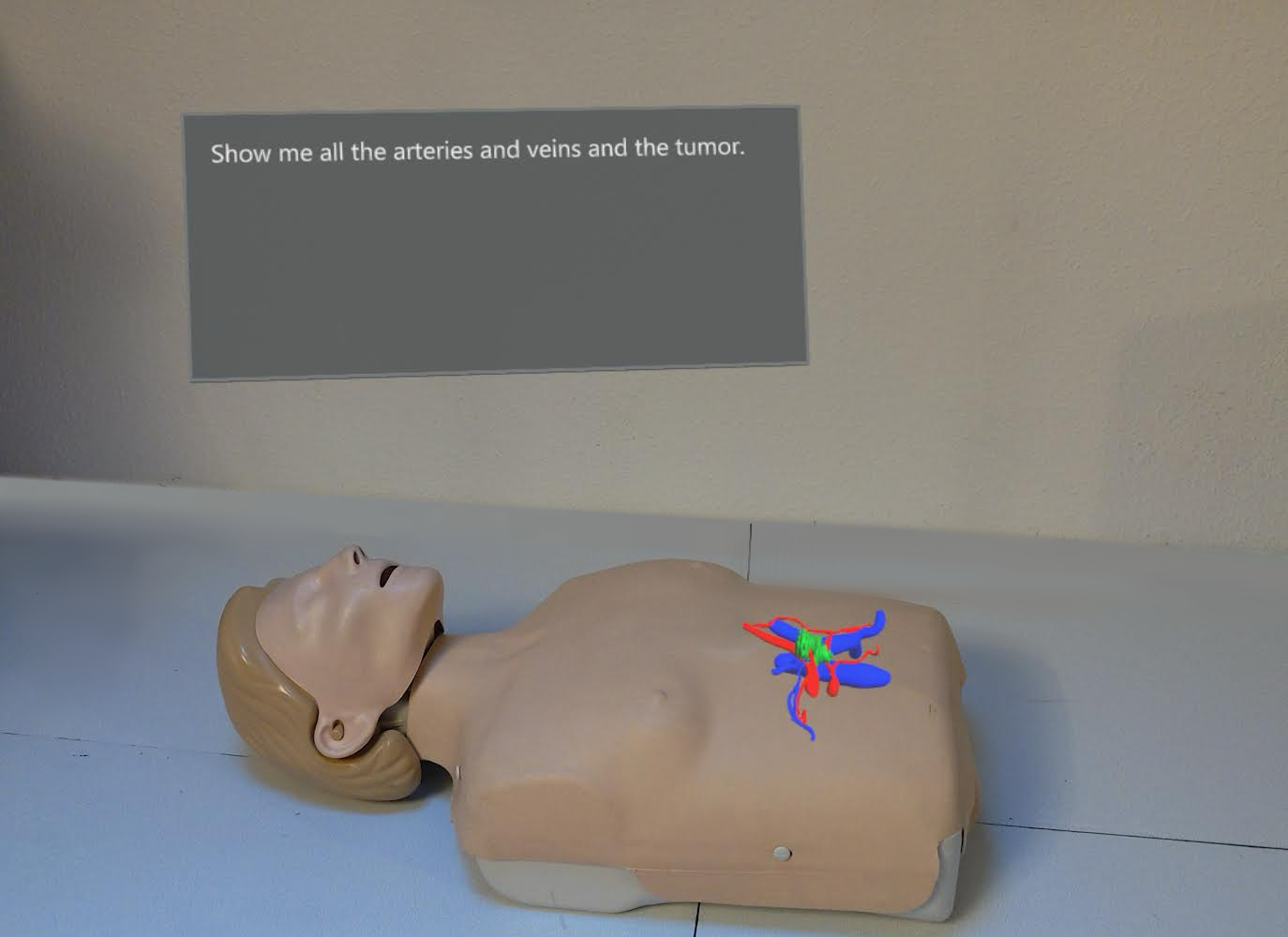}
    \caption{A captured image from the AR surgical assistance system using LLM-based VCUI. The manikin and transcription panel were used in the first study to simulate the visualization of the overlayed patient-specific 3D model during the surgical session.}
    \label{fig:manikin}
\end{figure}

During this study, we used a medical manikin to simulate the surgical scenario where the AR assistance system (\Cref{fig:manikin}). We used the reconstructed 3D models of two real patients with complex pancreas tumor localization with vascular involvement to efficiently address all the above-mentioned tasks. Both study groups included both patient cases. The participants were asked to stand around the table where the manikin was placed and position themselves as lead surgeon position (\Cref{fig:surgeryTable}, surgeon number 1) as the main decisions during the surgery and the above-mentioned tasks are usually performed and are decided by the lead surgeon. 

\subsection{Measures and Data Analysis}
We used quantitative and qualitative measures to evaluate different VCUIs. TCT (measured in seconds) was recorded for each performed task. We also recorded the attempt count for the successful completion of each task.

We used the SUS \cite{brooke1996sus} and NASA-RTLX \cite{hart2006nasa} questionnaires to evaluate system usability and perceived cognitive workload after the completion of all tasks using each VC method. We concluded with a semi-structured interview with each participant to gather qualitative insights about their experiences with each method.

To analyze the quantitative data, we formulate the following hypotheses:

\vspace{10pt}
\begin{frame}{}
\begin{minipage}{\linewidth}
  \myboxHT{\textbf{Hypotheses (Hs):}}{
  \normalsize
  \begin{enumerate}
     \item The LLM-based VC leads to lower task completion times.
    \item The LLM-based VC leads to lower cognitive load as measured by NASA-RTLX.
    \item The LLM-based VC has better usability as measured by SUS.
    \end{enumerate}

}
\end{minipage}
    
\end{frame}  
\vspace{10pt}

After confirming the normality of the data, we thus conducted one-tailed paired t-test to confirm or reject our hypotheses. For H3, we conducted a one-tailed Wilcoxon signed rank test instead, since data normality was violated.

All interviews were transcribed verbatim. Given the volume of the data, we followed the pragmatic approach to qualitative analysis as recommended by Blandford \textit{et al.} \cite{blandford2016qualitative}. Initially, two researchers analyzed 25\% portion of the data. Following this, we created a preliminary coding framework through iterative discussions. The remainder of the interview data was then divided equally among the two researchers for coding. In a concluding discussion, we refined the coding framework further and identified overarching themes.

\subsection{Procedure}
The study procedure started with participants completing demographic questionnaires to provide essential background information. To mitigate order bias and potential learning effects, all participants engaged with both studied VCUIs in a counterbalanced order with a different case for each VCUI. 
Prior to task execution, participants were familiarized with the AR system and its functionalities using each VCUI. Data recording was initiated after participants confirmed their ability to successfully interact with the system. Each participant systematically performed all experiment tasks for two sessions, each session using a different VCUI and different patient case. Task progression was subject to verbal confirmation of the participant regarding the accurate visualization of structures. Upon completion of the six tasks in each session, participants were prompted to fill out paper-printed questionnaires asking them to specifically answer the questions considering the experienced VCUI but not the patient case. After completing both sessions, a brief semi-structured interview was conducted to gather insights about participants' opinions regarding each VCUI. The whole study took approximately 40 minutes per participant. The study received the approval of the [removed for review] ethical committee board, ensuring that all aspects of the research adhered to established ethical guidelines. 

\subsection{Participants}
Our study included nine volunteered experienced surgeons with a mean age of 42.44 (SD = 7.49) and 14.33 (SD = 7.35) years of average surgical experience. We compared the participant number with the participant number required for a usability evaluation. The number of participants in this study falls above the acceptable range of $4 \pm 1$  \cite{caine2016local, hwang2010number}, considering their high expertise in the field. The participants reported an average of 11 to 50 times experience with AR technology and 2 to 10 times experience with LLM or generative AI systems including chatbots and conversational assistance systems. The detailed characteristics of the participants are given in \Cref{tab:char}.

\begin{table}[htbp]
\centering
\caption{Participant Characteristics (N=9). Likert scale values range from 1 to 5, with 1 being the least and 5 being the most frequent or proficient. SD = Standard deviation}
%\resizebox{0.4\textwidth}{!}{%
\begin{tabular}{lcc}
\toprule
Characteristic & Mean & SD \\
\midrule
Age (years) & 42.44 & 7.49 \\
Surgical Experience (years) & 14.33 & 7.35 \\
How many times used AR (1-5 Likert Scale) & 3.00 & 1.50 \\
How many times used LLM (1-5 Likert Scale) & 2.78 & 1.20 \\
English Proficiency (1-5 Likert Scale) & 3.89 & 0.60 \\
\bottomrule
\end{tabular}
%}
\label{tab:char}
\end{table}

% \begin{figure*}
%     \centering
%     \includegraphics[width=\textwidth]{Figures/TCT_combined_tasks.pdf}
%     \caption{The task completion times for each task using a different interaction method. Significant differences are marked with * for $p<0.05$ and with ** for $p<0.01$ }
%     \label{fig:TCT}
% \end{figure*}

\begin{figure*}[t]
    \centering
    
    \includegraphics[width=0.4\columnwidth]{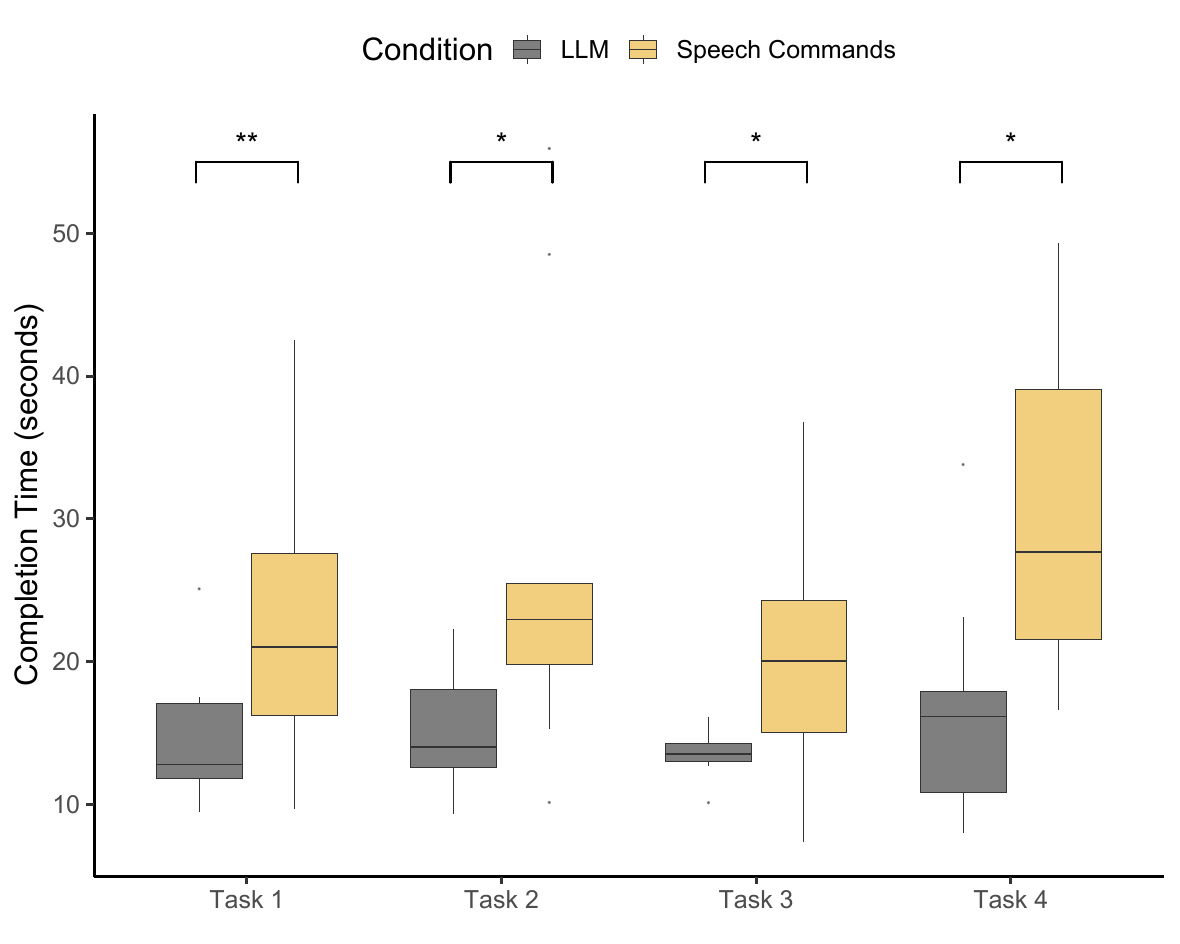}
    \hspace{1cm}
    \includegraphics[width=0.2\columnwidth]{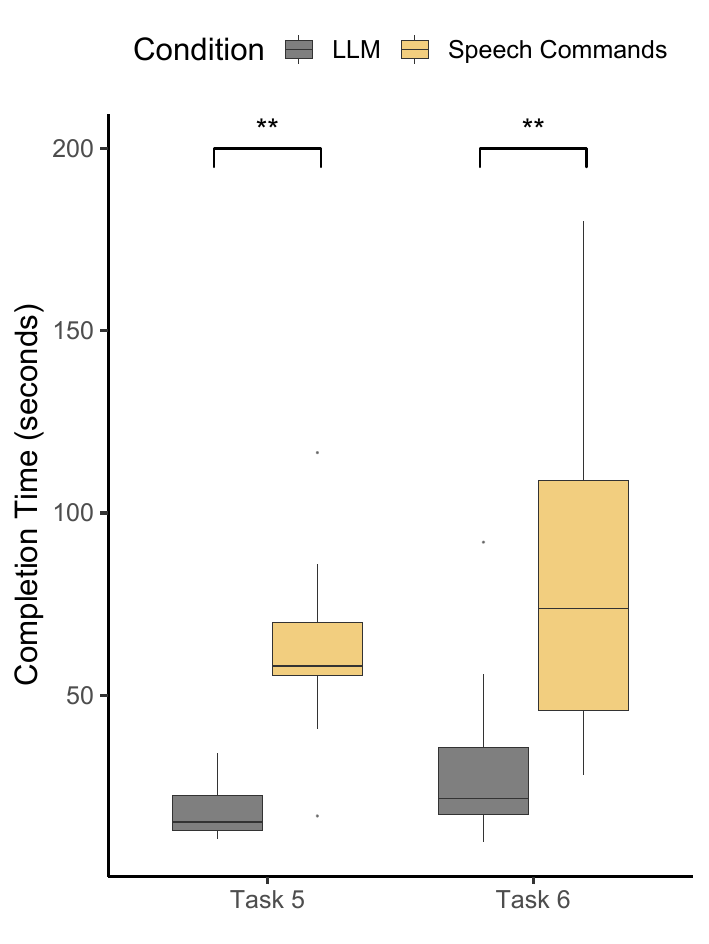}
    
    \caption{Task completion times for each task given the different VC methods (LLM, Speech commands). LLM yielded significantly lower completion times for all tasks (marked with * for $p<.05$ and with ** for $p<.01$).}
    \label{fig:TCT}
\end{figure*}

\subsection{Results}
\label{sec:study1Results}
\subsubsection{Quantitative Measures: TCT, NASA-RTLX, SUS}
The participants completed all six tasks using both methods: LLM, with a mean attempt count of 1.074 (SD = 0.328), and speech commands, with a mean attempt count of 1.370 (SD = 0.784) for successful completion. As shown in \Cref{fig:TCT}, individual one-tailed paired t-tests revealed that the LLM-based VC method yielded a significantly lower TCT for all tasks: Task 1 ($t(8)=-3.04$, $p < .01$), Task 2 ($t(8)=-2.33$, $p < .05$), Task 3 ($t(8)=-2.60$, $p < .05$), Task 4 ($t(8)=-2.56$, $p < .05$), Task 5 ($t(8)=-4.34$, $p < .01$), and Task 6 ($t(8)=-4.16$, $p < .01$). This result confirms H1.

\begin{figure}[htbp]
    \centering
    \includegraphics[width=2.5cm]{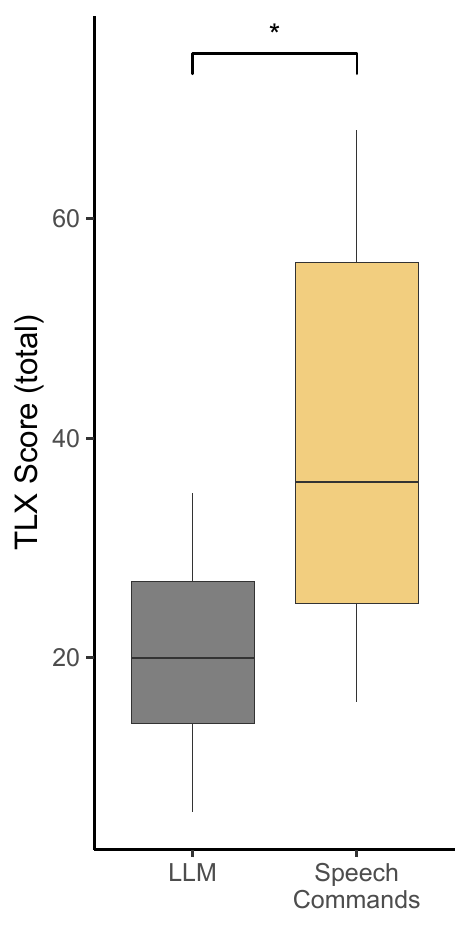}    \hspace{1cm}    \includegraphics[width=6cm]{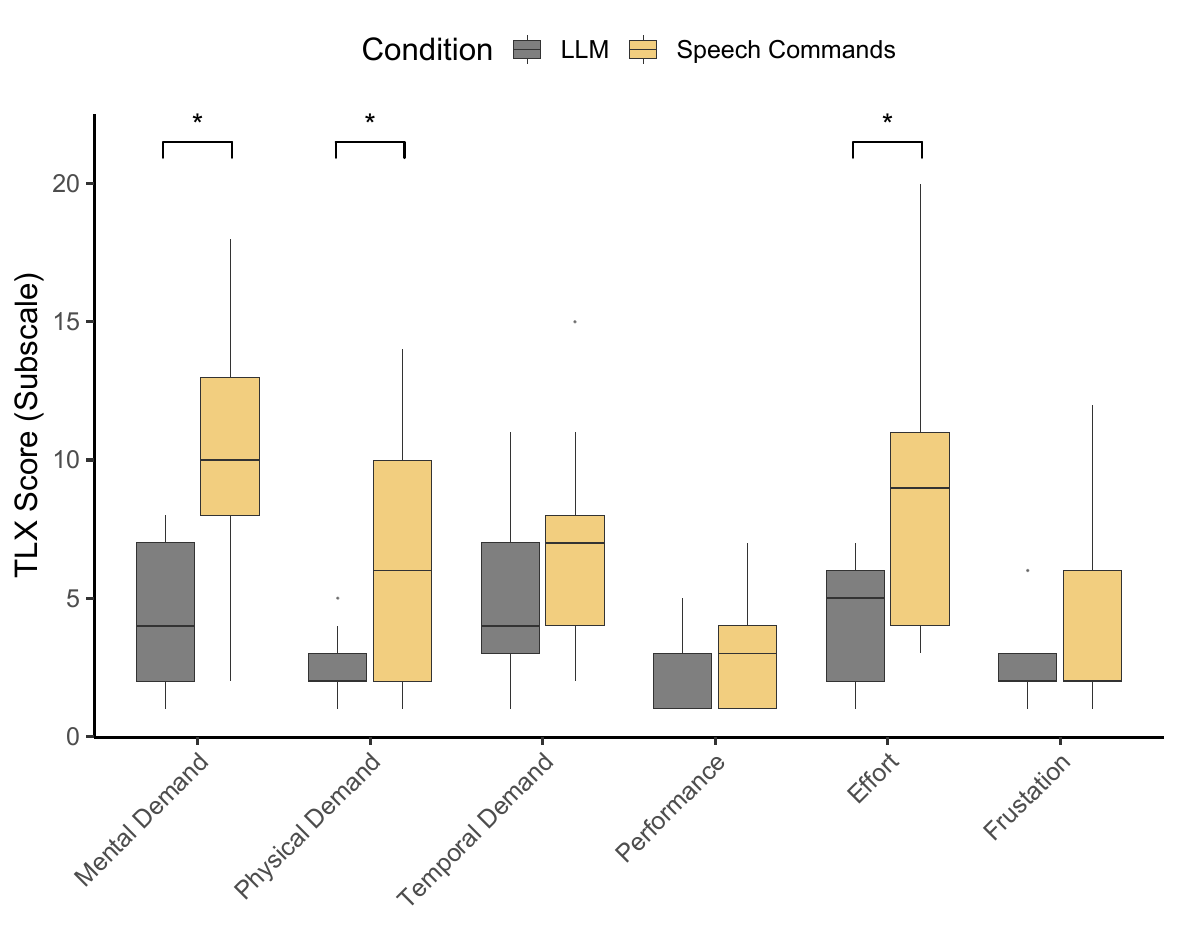}
    \caption{Total NASA-RTLX score (left) and individual subscale scores (right) given both VC methods (LLM, Speech commands). LLM yielded significantly lower scores (marked with *).}
    \label{fig:tlx}
\end{figure}

The overall NASA-RTLX score was significantly lower for the LLM-based approach ($t(8)=-2.24$, $p < .05$), confirming H2. Specifically, it scored significantly lower for the subscales mental demand ($t(8)=-2.71$, $p < .05$), physical demand ($t(8)=-2.35$, $p < .05$), and effort ($t(8)=-2.36$, $p < .05$). We found no significantly lower score for LLM-based VCUI for the other subscales, temporal demand, performance, and frustation. A visualization of this result can be found in \Cref{fig:tlx}.

% The results showed a significant difference ($p < 0.05$) %($p = 0.028$) 
% in the overall RTLX score between each interaction method (\Cref{fig:tlx}). Specifically, LLM performed significantly better in mental demand ($p < 0.05$) %$($p = 0.013$)
% , physical demand ($p < 0.05$)% ($p = 0.023$)
% , and effort ($p < 0.05$)%($p = 0.023$)
% categories compared to speech commands. However, no significant difference was found in temporal demand ($p > 0.05$)%($p = 0.198$)
% , performance ($p > 0.05$)%($p = 0.244$)
% , and frustration ($p > 0.05$) %($p = 0.125$)
% (\Cref{fig:tlx_seperate}).

Using Bangor et al.'s rating scale \cite{bangor2009determining}, the LLM-based VCUI was classified as "excellent" with a SUS score of 87.78, and VCUI using speech command was classified as "good" with and 79.17 SUS score. Despite the better performance of LLM, it did not yield a statistically significant higher SUS score. Thus, H3 could not be confirmed.

% \begin{figure}
%     \centering
%     \includegraphics[width=.5\columnwidth]{Figures/sus.pdf}
%     \caption{TODO: sus n.s.,  labels should be changed to LLM vs. Speech commands}
%     \label{fig:sus}
% \end{figure}

% \begin{figure}[h]
%     \centering
%     \includegraphics[width=.5\columnwidth]{Figures/tlx.pdf}
%     \caption{NASA-RTLX score given both interaction methods (LLM, Speech commands). LLM yielded a significantly lower score (marked with *).}
%     \label{fig:tlx}
% \end{figure}

% \begin{figure}[h]
%     \centering
%     \includegraphics[width=\columnwidth]{Figures/tlx_seperate.pdf}
%     \caption{Scores for individual NASA-RTLX subscales given both interaction methods (LLM, Speech commands). LLM yielded significantly lower scores for physical demand, mental demand, and effort (marked with *).}
%     \label{fig:tlx_seperate}
% \end{figure}

\subsubsection{Interview Findings}
 Our analysis identified three main themes: User preferences and experience, context of use, and limitations and future improvements

\textbf{User Preferences and Experience: }
Participants generally expressed a strong preference for the LLM VCUI over traditional speech commands. They appreciated the flexibility and intuitiveness of the LLM, which allowed for more natural communication and seemed to reduce stress by accommodating various phrasings and intents without requiring specific keywords. 

One surgeon reflected on the ability to articulate complex queries and receive accurate, contextually relevant information was particularly valued in high-stakes environments like operating rooms, saying: 

\begin{quoting}
    I personally think it's much less stressful to have such language support because I can say whatever I want, how I ever want to phrase it and the system realizes what I want. (P1)
\end{quoting}

Furthermore, they emphasized that not only the LLM-based VCUI provided a more natural way of communicating with the system but also helped them to reduce the burden of thinking and making decisions about the requirements of the task by performing context-aware function calls. P2 pointed out this aspect stating:

\begin{quoting}
    %With the first method, with these voice commands, it wasn't so difficult. But then when I used this large language model, I realized that you have to think about the first method and then find it. And then you have to give the voice commands. 
    For example, for tumor infiltrations, you have to first look at the tumor and which vascular system is infiltrated, then tell each time what to open or turn on [Speech Commands]. That's why I think it was much better with LLM, with large language model, because I simply asked and that showed. (P2)
\end{quoting}

\textbf{Context of Use: }
The interviews revealed that, while both VCUIs are usable, the specific context and situations in which the system is employed may highlight the unique potential of each method. While the LLM-based method could be highly beneficial in stressful and time-critical conditions, the voice command might be a better option for surgeries that do not have time criticality. P8 and P9 highlighted this by following statements:

\begin{quoting}
 It depends on which time I have to use. So for example, if you say there's an emergency, see, then I would prefer the AI [LLM].  
 In obesity surgery I have time. I have no emergency. So I don’t have tumor and I have more time. In the process of the operation you can say remove this, remove that. So step by step [Speech Comands]. (P9)
\end{quoting}

\begin{quoting}
If the patient bleeds, even a few seconds wait is already annoying (...) nevertheless, even with bleeding I find the second one better [LLM], because you can say directly what you want to see without thinking. (P8) 
\end{quoting}

\textbf{Limitations and Future Improvements: }
The interviews revealed the importance of accurate dictation and robust speech recognition service for both VCUIs. The visual feedback on the real-time transcription of the user query in the LLM-based method caused extra confusion, showing the potential misjudgment about the system's capability by the user. As users attempted to correct misinterpreted words upon observing incorrect transcriptions, the clarity of their requests diminished, leading to decreased performance in triggering the relevant functions by LLM. Conversely, the LLM would typically mitigate such errors by interpreting them as typographical mistakes, thereby maintaining higher accuracy in understanding and responding to user commands. P4 pointed out this matter by raising attention to the system's capability being affected by misinterpreted words due to the different pronunciations saying: 

\begin{quoting}
     The less I say, it's supposed to be easier for the system to understand me, right? I don't know. I was thinking when I say too much then the system doesn't understand me because I said too much. With shorter words you minimize the misunderstanding when someone pronounces it differently. (P4)
\end{quoting}

The surgeons also raised concerns about the default listening time that was set in the LLM-based approach. P3 suggested adopting an approach where the initialization and ending of the listening time could be activated by some keywords to refrain from waiting if the query time is less or longer than the default listening time, saying:

\begin{quoting}
    For example, you say assistant or something to begin. But can I also say end of sentence or so that I don't have to wait those couple of seconds.~(P3)
\end{quoting}

Additionally, participants mentioned that the LLM-based VCUI required a clear statement of the request. Despite the ease of use, the interaction would be even easier over time as one would learn how to clearly phrase their request. P3 reflected on this saying:

\begin{quoting}
    When I tried the second one [LLM], with the AI, I think it's even easier to use if you've done it before. Then you have the routine of what you have to say so that the device understands what I want, and then it's easier and quicker. I mean how I should formulate my question so that the system understands me and shows the result that I want. (P3)
\end{quoting}

%Dictation errors-> completely affect speech command functionality, but LLM still survives even with errors caused by dictation service mistranscriptions?

\subsection{Implications}
Our study comparing LLM-based VCUI to a VCUI using speech commands has revealed implications for the integration of such technologies into surgical settings. These implications highlight the potential benefits and necessary refinements for practical application:

\textbf{Potential superiority of LLM-based approach in critical surgical moments:}
The LLM-based approach demonstrated advantages, particularly when a decision-making situation was involved. It exhibited significantly reduced execution times across various tasks. As tasks increased in cognitive demand, particularly in Task 5 and Task 6, the disparity in execution times became more pronounced, reflecting the challenge of mental workload and decision-making when using speech commands. Moreover, assessments of cognitive load indicated a lower mental demand with the LLM-based approach. This convergence suggests that LLM-based VCUI could offer a superior option in real surgical environments, where timely decisions are required during constrained time frames.

\textbf{Need for system refinements prior to real-surgery evaluation:}
However, our findings also illuminate areas necessitating refinement before practical deployment in surgical settings. The real-time transcription panel introduced confusion as users attempted to rectify misinterpreted words, compromising the clarity of sentence context. Additionally, while LLM-based VCUI facilitated quicker task completion, further enhancements in dictation service are essential to mitigate any remaining delays and optimize task execution times.

\section{Case Study: Evaluation During Actual Surgery}
\label{sec:study2}
\begin{figure}[htbp]
    \centering   \includegraphics[width=0.55\columnwidth]{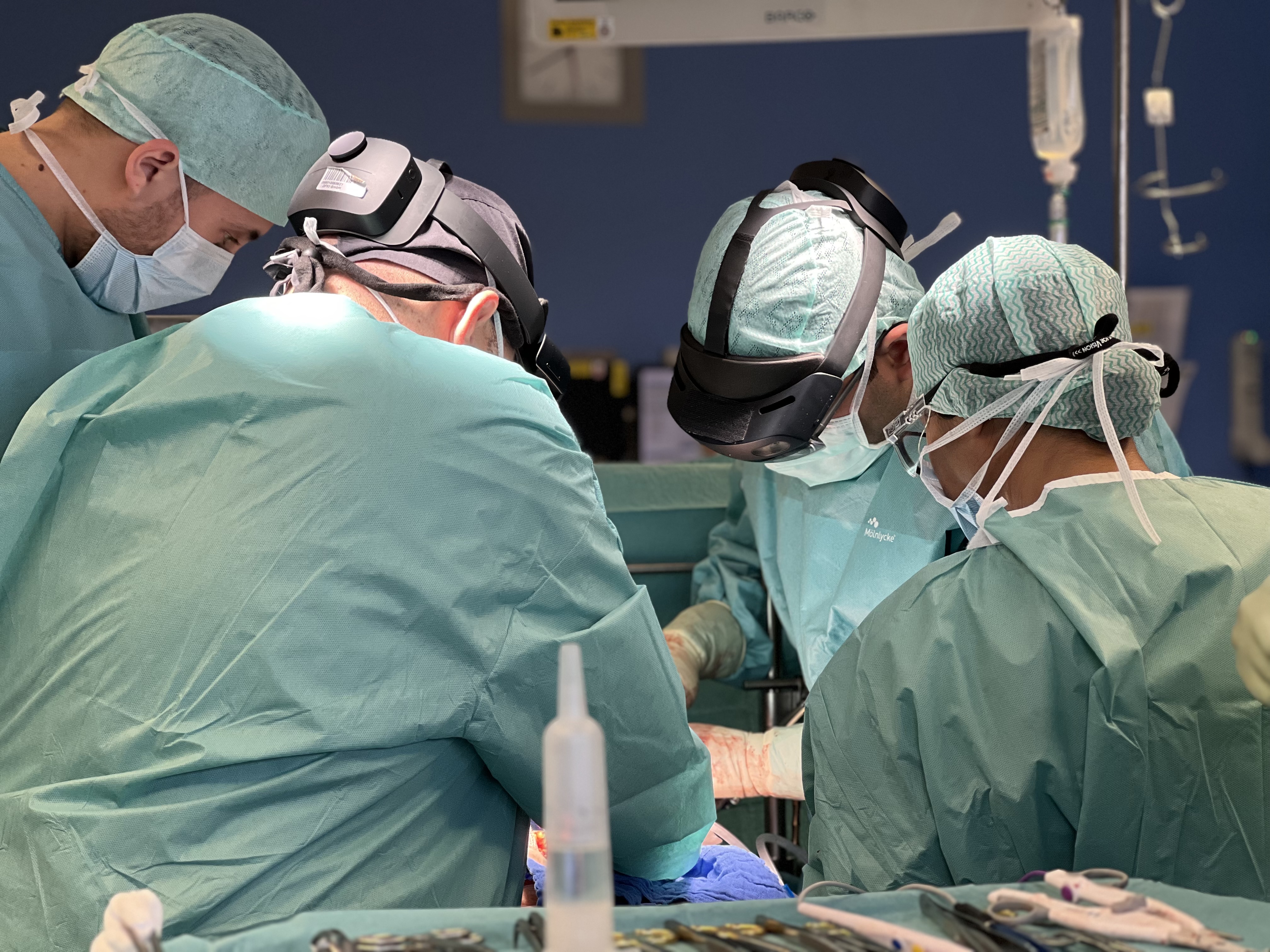}
    \hspace{0.1cm}
     \includegraphics[width=0.38\columnwidth]{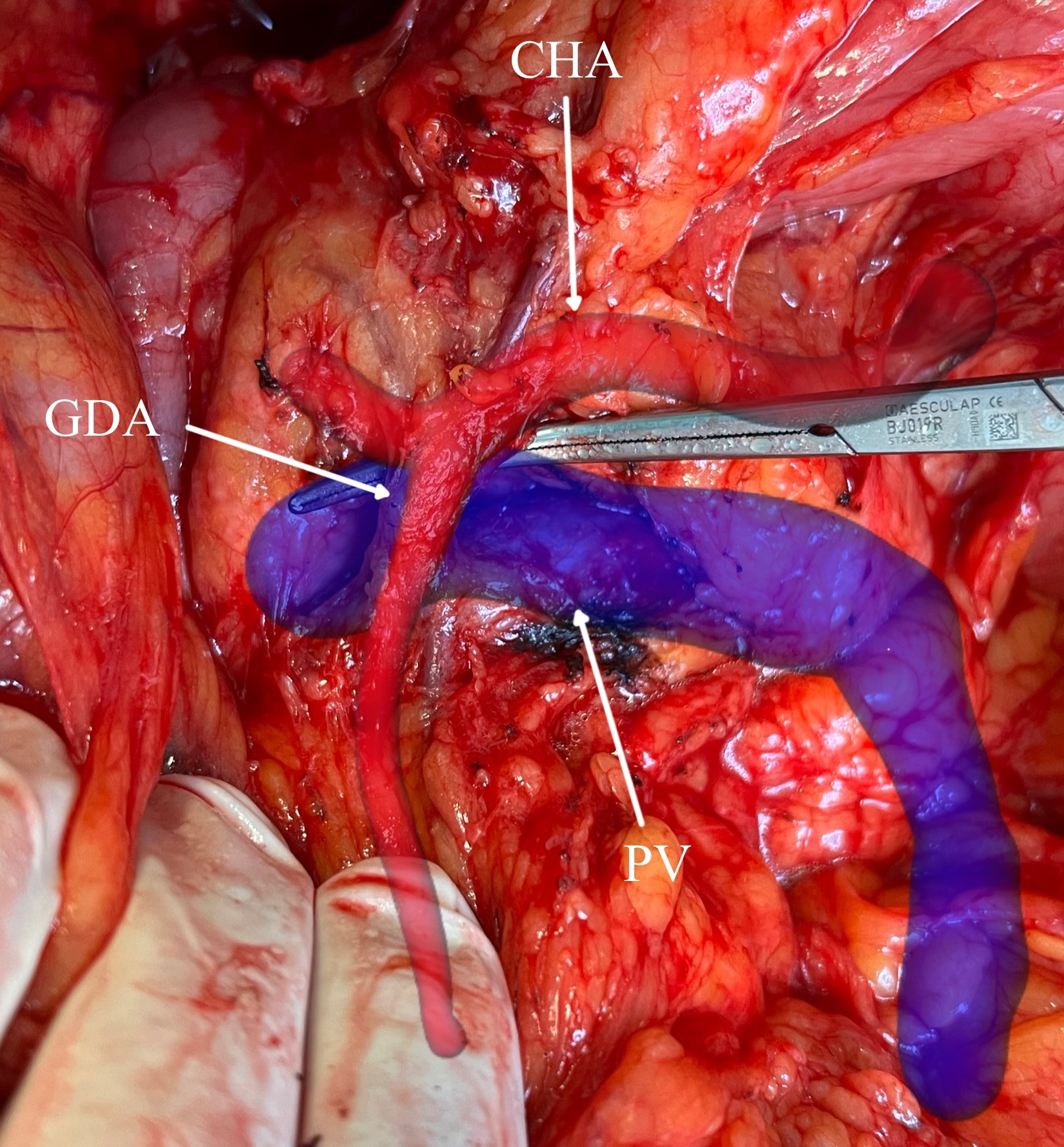}
    \caption{Actual pancreatic surgery. The right picture shows a snippet from the application view captured from a surgeon's device. GDA: Gastroduodenal Artery, PV: Portal Vein, CHA: Common Hepatic Artery}
    \label{fig:surgery}
\end{figure}

Following the proven usability of both VCUI systems in our initial user study (\cref{sec: study1}), this study aimed to further evaluate these VCUIs in a real surgical setup, addressing our third research question (RQ3). This phase sought to confirm our findings under actual surgical conditions, which can differ significantly from laboratory environments due to factors such as higher stress levels and time constraints. Building upon our findings from our first user study (\Cref{sec:study1Results}), we have first performed refinements (\Cref{fig:procedure}, Refinement) to our approach and later evaluated each VCUI during a pancreatic tumor removal surgery (\Cref{fig:surgery}).

We performed the following refinements to our LLM-based VCUI: The transcription panel providing real-time feedback to visualize the transcription of the voice query panel was removed, as it was observed to cause more confusion. Users attempting to correct what they perceived as misinterpreted words during their query can diminish the efficiency of the LLM method, as the context of the sentence may become unclear. Instead, we used conversational audio feedback similar to those commercially available conversational assistants such as Siri \footnote{Siri is Apple's voice-activated virtual assistant, available on iOS devices such as iPhones and iPads.}. We used sound saying "OK" to indicate receiving the user query and "Please state your request differently" when the LLM response did not yield any of the defined system functions. Furthermore, we adapted the listening time after activation of the dictation service. The user request would automatically send to the LLM model after one and half seconds of silence without waiting for any further default time.  

After performing refinements, we deployed our previously developed AR Assistance system designed to visualize 3D model of the patient during the surgery with the capability of both VCUIs. 

\subsection{Study Design}
To evaluate our LLM-based VCUI and compare its outcomes, such as TCT and cognitive load, with the speech command during actual surgery, we exclusively employed qualitative measures and conducted post-operation interviews. This decision was driven by the inherent variability in each patient case, which might inevitably affect cognitive load measurements due to the unique nature of each surgical procedure. Similarly, task execution time would be influenced by the specifics of the surgery being performed. Consequently, a direct comparison between the two methods across different surgeries would not yield meaningful results. Thus, we chose to gather insights through interviews and observations, with an observer researcher participating in the surgery sessions, making notes and observations about user interaction with the system. This approach allowed us to assess the impacts of each VCUI, enabling us to collect qualitative data that could inform future improvements and implementations.

\subsection{Procedure}
To validate our approach in an ecologically valid environment, we conducted clinical trials involving the intraoperative evaluation in patients with underlying (borderline) resectable pancreatic tumors who required various types of pancreatic resection. The trials took place at [removed for anonymized review], certified to perform pancreatic tumor resection surgeries.

The study protocol received approval from the Ethics Committee of the Medical Association of %Saarland 
[removed for anonymized review] under registration number: [removed for anonymized review]. %(registration number: 159/23).
%the Medical Association of Saarland (registration number: 159/23). 
The protocol of our study was also registered at ClinicalTrials.gov under the registration number: [removed for anonymized review]. %NCT06208579.
 
Patients provided informed consent prior to surgery. All participating surgeons were fully briefed on the experimental nature of the method and the device used. They voluntarily agreed to use the system during the surgeries, assuming full responsibility for its operation and the outcomes. The surgeons were also informed that they could discontinue the use of the system at any point if necessary, without obligation.

To avoid first-time use bias two of the surgeons (P1, P2) who participated in study 1 participated in this study. Each surgery began with two surgeons equipped with our designed wearable assistance system with both VC method capabilities.
In each session, surgeons were asked to use only one of the VC methods. However, they always had the choice to use the other method if it was essential for the course of the surgery. We conducted interviews with surgeons after each surgery session about their experience with each VCUI. 

\subsection{Results}
No technical difficulties regarding the VCUIs were observed during both sessions and both surgeons used the system with the assigned VCUI throughout the surgery. 

Interviews with two surgeons who experimented with both VCUI across two surgical procedures proved the feasibility of our LLM-based VCUI in a real surgical environment in line with findings from our first study.

The LLM's capability to discern user context and analyze patient data facilitated the surgeons in adjusting the visualization of patient 3D models according to the tumor's proximity more efficiently specifically in the initial preparation phase of the operation. This benefit was encapsulated by a participating surgeon, who remarked:

\begin{quoting}
    Today's patient had an anatomical anomaly so we had to be more careful identifying the vessels during preparation so we don't damage them because they were so close to the tumor. So we had to change the visualization a lot. At that moment actually the LLM was a big help actually because it saved us a lot of times. (S2)
\end{quoting}

S1 also reflected on this topic saying: 

\begin{quoting}
    I think the biggest difference between two [VCUIs] was during the initial phase of the operation where we usually use the system more to identify the vessels. But when the vessels are identified and already visible we don't interact with the system much. (S1)
\end{quoting}

Unlike the speech command method which requires precise pronunciation of predefined keywords, the LLM system maintained a more natural communication flow. This aspect was profoundly appreciated, as S2 shared:

\begin{quoting}
    Voice commands also worked fine but the issue with the voice commands is sometimes when people are talking around the table you have to say a word 100 times until the system detects it. LLM is more forgiving if that's a correct word to use. (S2)
\end{quoting}

Additionally, S1 reported on the benefits of using natural communication to control the system and also sharing information with other staff around the table. S1 reported: 

\begin{quoting}
    When I say a single word usually other staff surgeons don't know what I am doing because they don't see what I see in the device. But when I talk to the system the way how I talk normally, then they know, ok, now I am trying to see where some vessels are when I say, for example, show me the vessels near the tumor or like I want to see mesenteric artery. (S1)
\end{quoting}

S2 also highlighted the benefits of the performed refinements, noting the reduced confusion from removing the transcription panel and improved system response times: 

\begin{quoting}
    This time with LLM system we didn't have to wait much for the system to react so it was way better and less annoying. Also, I think removing the panel was a good decision as I didn't see what the system understands so I didn't worry much about correcting my request and the system worked even better. (S2)
\end{quoting}

\section{Discussion}
The introduction of AR-based surgical assistance systems has significantly transformed surgical practices, offering an enhanced level of precision and support. As these technologies evolve, the choice of interaction method to control the system becomes a pivotal consideration. Our study introduces a novel VCUI for surgical ARAS using speech recognition and LLM and conducts a comparative analysis with the conventional VCUI using speech recognition and speech commands, focusing on enhancing operational efficiency and user experience in the critical context of surgery. Importantly, we tested both methods in controlled laboratory settings and real surgical environments, offering a robust evaluation of their practical application and performance. This dual-context approach allowed us to gather comprehensive insights into the efficiency, cognitive load, user preferences, limitations, and situational applicability of each VC method. 

\subsection{Efficiency and Cognitive Load: LLMs versus Speech Commands}
The SUS scores, along with the successful implementation of both VC methods in simulated and real surgical environments, demonstrate their usability and confirm their applicability in critical medical settings. However, the distinction in performance, especially in time-sensitive scenarios like the initial phases of surgical intervention, which is the most mentally demanding phase, underscores the criticality of choosing the right control and interaction method.

The use of LLMs significantly outperformed traditional speech commands in TCT. This efficiency is attributable to the LLMs' ability to generate context-aware outputs and execute multiple functions simultaneously to achieve a certain undefined functionality, a quality unattainable with keyword-specific speech commands without implementing further keywords to perform this task. As functionalities expand, the speech command method suffers from scalability issues, requiring an ever-increasing list of keywords. Conversely, LLMs streamline this process, enabling parallel function execution based on user requests without necessitating an extensive set of unique commands. 

A more intriguing aspect of LLMs lies not just in determining which function to call based on user requests, but also in acting as an intelligent assistant and generating outputs which normally requires a complex decision-making process. This was particularly evident when LLM successfully generated the correct output to call certain system functionalities even when the function name or its specific purpose was not directly mentioned in the user’s query. The capability of LLMs to generate context-aware outputs using all available information represents a significant advancement towards truly intelligent user interfaces and assistance systems.

A detailed analysis of user interaction logs with the LLM revealed its ability to successfully make decisions in numerous instances where users would otherwise have had to decide themselves. This difference in performance compared to speech commands was particularly evident. Despite the system functionalities being simple and identical in both cases, the reasons for employing these functionalities were often complex. With speech commands, the user (a surgeon) needed to decide and then instruct the system to make specific changes in visualizations using keywords. In contrast, the LLM-based system handled the decision-making process.

For example, in task 6, P2 asked the LLM, "Can you show me what should be resected?" Despite no specific information or annotation regarding the task description or the structures to be enabled in this context, the LLM correctly decided to display the tumor and the infiltrated veins and arteries within the resection margins by invoking multiple system functions simultaneously. This decision-making process is highly complex, relying on factors such as patient history, tumor position, and surgical guidelines regarding resection margins.

This feature demonstrated by the LLM not only reduced task completion times but also significantly diminished cognitive load. This was also evidenced by the lower scores in the NASA-RTLX, indicating a more intuitive and less burdensome interaction for the user—something unachievable with speech command methods without pre-implementing more complex functions into the system.

In applications such as surgical navigation systems, where system interaction is part of a decision-making process contingent on the patient case and scenario, pre-implementing an all-encompassing solution is very challenging. On the other hand, other specialized voice assistance methods, such as those using machine learning \cite{gowthamy2023enhanced}, require training 
the system with specific user terminology for different scenarios. Additionally, they come with significant processing costs that affect the performance of wearable devices.
Here, LLMs can provide significant benefits, offering a more adaptive, context-aware, and efficient approach to managing complex tasks even though the system itself remains simple regarding the functionalities.

%Moreover, our endeavor to find a feasible and optimized interaction approach revealed the potential of LLMs to not only facilitate function execution but also to aid in decision-making processes. We discovered with enough data and example provided, the LLM could make context aware decisions and perform complicated tasks by calling a mixture of simple functions without need to preimplement more complicated functions in the system itself.
By analyzing patient-specific data and examples given in the dynamically generated initial prompt, LLMs can offer custom recommendations, enhancing the support system's utility in high-pressure situations. This capability to interpret context and call relevant function combinations offers surgeons a richer, more contextual understanding of the patient's data, including visualization of details in the 3D models, an attribution that with conventional voice assistant systems using speech commands cannot be achieved \cite{gowthamy2023enhanced}.

Despite the apparent advantages of the LLM method in facilitating quicker, multi-functional requests, our study also highlighted a perception mismatch among some participants. They perceived speech command execution as a faster method for task completion, despite objective evidence showing the LLM method reduced TCTs. This discrepancy may be linked to the system's default listening time (minimum ten seconds or wait for 2 seconds of silence if longer than ten seconds) following the user query. It suggests the necessity for an adaptive approach in managing the activation and deactivation of the system's listening duration for the LLM-based method to ensure the receipt of full user queries without a long wait. As a shorter listening period could prematurely send incomplete queries to the LLM, while a longer period might unnecessarily delay the system's response. 

\subsection{Pros and Cons: Balancing Control and Transparency}

In our study we found out that, speech commands, with their direct and deterministic nature, afford users a clear understanding of cause and effect. This transparency in interaction fosters a sense of reliability and control, essential in high-stakes environments like surgery. However, this method's scalability and flexibility are constrained by the need to predefine every command, which can limit the system's responsiveness to complex or unforeseen requests.

On the other hand, LLMs represent a paradigm shift towards more fluid, conversational interactions. By understanding and processing natural language, these models offer a dynamic and flexible interface that can interpret a broad spectrum of user requests. However, this sophistication comes with a degree of opacity. The "black box" nature of LLMs can obscure the pathway from request to action, potentially undermining user confidence if the system's reasoning and decision-making processes are not sufficiently transparent.

Our study revealed that the LLM system could effectively compensate for errors in the dictation and speech recognition system. Unlike speech commands, which necessitate precise pronunciation, the LLM system can infer the user's intent by analyzing the context of the query rather than focusing on individual words. This capability significantly enhances the system's flexibility and user-friendliness.

However, it became evident that providing users with real-time visual feedback of transcription could inadvertently lead to misjudgments about the system's capabilities. Users attempting to correct what they perceive as misinterpreted words during their query can diminish the LLM method's efficiency, as the context of the sentence may become obscured. This observation underscores the critical need for designing user feedback mechanisms that do not compromise the clarity of communication or the efficiency of the system control method.

Furthermore, mitigating these challenges necessitates clear communication about the system's operational boundaries and capabilities. Users need to understand not just how to interact with the system, but also the underlying principles guiding its responses. This understanding is crucial for formulating effective requests, especially with LLMs, where the context and specificity of language can dramatically influence outcomes. Training and educational programs play a pivotal role in this regard, equipping users with the knowledge to navigate the system's complexities and leverage its full potential.

We believe that a hybrid approach, integrating both speech commands and LLM capabilities, emerges as a promising solution to balance control with transparency. By allowing users to switch between modes based on the task's complexity or urgency, such a system combines the directness of speech commands with the adaptability of LLMs. For routine tasks or when precision is paramount, predefined speech commands could offer the most efficient pathway. Conversely, for tasks that require time consuming decision-making process or when additional context is required, the LLM recommendations could provide a faster solution.

Implementing a hybrid model also entails designing interfaces that intuitively signal which mode is in operation, thereby maintaining user awareness and trust. Visual or auditory cues could indicate the system's current state, whether executing a direct command or processing a more complex LLM-based request. Moreover, offering users the ability to override or specify the control mode empowers them to use the system's capabilities depending to their immediate needs and preferences. By carefully navigating the trade-offs between control and transparency, and by fostering an environment of continuous learning and adaptation, we can develop systems that not only enhance surgical outcomes but also align with the users' operational and cognitive needs.
\subsection{Ethical Considerations}
Maintaining ethical standards is crucial for preserving trust in medical research and innovation. By adhering to ethical guidelines, researchers and practitioners demonstrate their commitment to prioritizing patient safety and well-being over technological advancements.

In this study, we emphasized ethical adherence throughout all stages. We ensured that the introduction of the AR system did not compromise the safety of patients or surgeons, nor did it undermine the integrity of the surgical process.

To achieve this, we initiated the study only after obtaining full approval from the relevant medical ethics review board. All participants, including surgeons and patients, were thoroughly informed about the study, with their participation contingent upon a clear explanation and the collection of informed consent. Patients were made aware of the potential risks, benefits, and alternatives to ensure their participation was both voluntary and fully informed.

We also ensured that neither the AR system nor the VC method used did replace the surgeon’s judgment, maintaining human decision-making in surgical procedures, and surgeons retained complete control over the system, consistent with the Fundamental Principles of Ethics \cite{varkey2021principles}.

Our key takeouts from this study regarding ethical considerations for future studies are as follows:
The system should function solely as a supplementary tool to assist the surgeon without replacing the surgeon’s expertise.
The surgeon must retain ultimate decision-making authority, ensuring patient safety and adapting to the unique aspects of each case.
Human intuition and experience should remain the final safeguard in surgical procedures.

\subsection{Limitations and Future Work}
Even though the results of this study are promising steps towards using LLMs not only as a VCUI but also as intelligent assistants in the medical domain, our findings are limited to the specific functionalities of ARAS we used in this study. While these functionalities are integral to such applications, a broader understanding of LLM capabilities requires further research. This should involve more complex system functionalities and tasks to fully explore and validate the potential of LLMs in diverse and demanding scenarios. In our future work, we intend to broaden the scope of our investigation into the capabilities and opportunities presented by LLMs in surgical assistance systems beyond our current application as a function caller and VC method. By leveraging the advanced natural language understanding and processing capabilities of LLMs, we hope to uncover new ways in which these models can contribute to the enhancement of surgical outcomes, efficiency, and safety.

\section{Conclusion}
Our comparative study of two VCUIs within an AR-based surgical assistance system highlights the distinct advantages and considerations associated with speech commands and LLM. We found that the LLM-based VCUI offered significant improvements in operational efficiency and reduced the cognitive load of users by allowing for natural, conversational interactions and the ability to generate context-aware system behavior by executing multiple functions concurrently. However, the choice between LLMs and speech commands is not clear-cut, despite higher preference towards LLM user preferences may vary based on perceived control, transparency, and the context in which the system is employed. While speech commands provide a sense of direct control and transparency, LLMs require clear instructions to function optimally, which can sometimes challenge users. The idea of a hybrid model emerges as a promising solution, aiming to combine the strengths of both approaches to cater to a broader range of needs and situations in surgical settings. Looking ahead, we plan to expand our exploration into the potential of LLMs as conversational assistants that not only could control the system but could participate more in the decision-making process, further enhancing the capabilities of surgical assistance systems. This study lays the groundwork for future advancements in surgical technology, emphasizing the importance of the involvement of end-users during design and evaluation and the need for systems that balance efficiency, cognitive ease, and adaptability to the fast-paced, complex nature of surgical environments.

\bibliographystyle{plain}
\bibliography{References.bib}


\appendix

\section{Initial Promt Json Format}

\begin{lstlisting}[breaklines,basicstyle=\scriptsize\ttfamily]

Initial_prompt = {
  "description": "Depending on given sentences, Return 
  only appropriate method or methods from the executable
  methods list without explanation.",
  "executableMethods": [
    "function_A(variables)", ..., "function_X(variables)"],
  "organTypes": [
    "Organ_A", ..., "Organ_X"],	
  "OrganCategories": [
    "Category_A", ..., "Category_X"],
  "distanceData": [
    " Organ_A": { " Organ_A": xx, ..., " Organ_x": xx}, 
    ...,
    " Organ_x": {" Organ_A": xx, ..., " Organ_x": xx}],
  "guidlines": [
    " rule_A":" description of rule_A,
    ...,
    " rule_x":" description of rule_x ],
 "sentencesAndResultsExamples": [
    { "sentence": "Show me all of the arteries",
      "result": "function_{xx} (variable_xx)"
    },
    ...,
    { "sentence": "Show me the infiltrated vessels",
      "result": {"function_{yy} (variable_y1, variable_y2)", 
      ..., "function_{yy} (variable_y1, variable_y2)" }}]
}
\end{lstlisting}

\label{appendix_json}

\end{document}